%Paper: astro-ph/9503103
%From: JCARLOS@vax.iagusp.usp.br (JOSE' CARLOS N. DE ARAUJO)
%Date: Tue, 28 Mar 1995 10:38:13 -0300
%Date (revised): Tue, 28 Mar 1995 16:35:15 -0300

\magnification=1000
\pretolerance=10000
\def\PN{\par\noindent}
\baselineskip 12 pt
\centerline {\bf Deformation of Rapidly Rotating Compact
Stars\footnote{$^*$}{\it Send offprint requests to: J.C.N. de Araujo
(Universidade de S\~ao Paulo)}}

\vskip 3 true cm
\centerline {J.C.N. de Araujo$^{1,2}$,
J.A. de Freitas Pacheco$^1$, M. Cattani$^3$ \& J.E. Horvath$^1$}

\vskip 18 pt
\centerline {$^1$Instituto Astron\^omico e Geof\'\i sico}
\centerline {Universidade de S\~ao Paulo, CP 9638 - 01065-970}
\centerline {S\~ao Paulo -  SP, Brasil}

\vskip 18 pt
\centerline {$^2$SISSA/ISAS}
\centerline {Strada Costiera 11, I-34014}
\centerline {Trieste, Italy}

\vskip 18 pt
\centerline {$^3$Instituto de F\'\i sica}
\centerline {Universidade de S\~ao Paulo CP 01498}
\centerline {S\~ao Paulo -  SP, Brasil}

\vskip 6 true cm
\baselineskip 18 pt
\par\noindent J.C.N. de Araujo et al.: Rapidly Rotating Compact Stars
\par\noindent Proposed Journal: Main Journal
\par\noindent Section: 6
\par\noindent Thesaurs code numbers: 02.04.1, 02.05.2, 02.18.8, 08.14.1,
08.16.6, 08.18.1
\par\noindent Author to whom proofs should be sent: J.C.N. de Araujo
(Universidade de S\~ao Paulo)

\vfill\eject
\baselineskip 18 pt

\PN {\bf Abstract.} We have developed a numerical code to study the
deformation ($\varepsilon = (I_{zz}-I_{xx})/I_{zz}$, where $I_{ii}$ are
the moments of inertia) of neutron stars in rapidly rotation in a fully
general relativistic calculation. We have found that the deformation
is larger, depending on the angular velocity, than is generally assumed for
gravitational wave estimations. Calculations were performed by employing
the Bethe-Johnson I EOS (equation of state) and a new set of models by
the Frankfurt group including $\Lambda$ hyperons for several choices of
their coupling constants to ordinary nucleons. Possible implications for
gravitational wave searches are briefly discussed.

\vskip 18pt


\vfill\eject

\PN {\bf 1. Introduction}

\par\noindent The problem of detection of gravitational waves is nowadays of
great interest. There are many experiments in development
to detect possible sources of gravitational waves.

\par From the theoretical point of view several
possible sources of gravitational waves have been advanced, like
close binary systems, non-spherical collapsing stars, non-radially
pulsating compact stars, wobbing pulsars, among others.
We shall address in this paper, the particular case
of the expected emission of gravitational waves from wobbling pulsars.

\par The dimensionless amplitude of the gravitational waves,
due to the wobbling neutron stars, depends on some poorly known
parameters like the so-called ``gravitational deformation''
$\varepsilon = (I_{zz} - I_{xx})/I_{zz}$ and the wobble angle $\theta_w$.
Concerning the ``gravitational deformation'' $\varepsilon$, values
in the literature of the order
of $\varepsilon \sim 10^{-3} - 10^{-4}$ are generally
assumed by different authors, e.g. Pines \& Shaham (1974), Zimmermann (1978),
Shapiro \& Teukolsky (1983), and Barone et al. (1988), to
be adequate values for
``slow'' radio pulsars. It is worth mentioning that, even though implicit
in the calculations, previous extensive studies on rapidly rotating neutron
stars like those by
Butterworth \& Ipser (1976; BI76), Friedman et al. (1986; FIP86),
Komatsu et al. (1989a, b; KEH89a, KEH89b), Eriguchi et al. (1994),
Lattimer et al. (1990; LPMY90), Cook et al. (1992, 1994a, 1994b;
CST92, CST94a, CST94b), Bonazzola et al. (1993; BGSM93) or
Salgado et al. (1994; SBGH94) have not addressed their results to
the analysis of such a relevant parameter for the emission of
gravitational waves.

\par The aim of this paper is to show  that the ``gravitational
deformation'' $\varepsilon$ (see appendix for details)
may be much larger than the values usually found
in the literature, namely, $\varepsilon \sim 0.1 - 0.2 $ at rotation
rates near the break-up value.
These values are 2-3 orders of magnitude
higher than those commonly adopted for the calculation of the gravitational
emission rate of wobbling neutron stars. This is an important point
since the actual pulsar population includes a large subpopulation for
which $\varepsilon$ due to rotation
would be found in the latter range, and this in
turn enhances the prospects for detection of wobble radiation
from the abundant galactic members, even assuming modest values
for the wobble angle.

\par It is important to note that, although we have performed a
fully general relativistic calculation of  $\varepsilon$,
the actual value
of this parameter might be lower than we have found.
This is due to the fact that an actual neutron star is much more
complicated than the simple fluid that we have assumed and
it should have, as it is generally accepted, a crust that behaves
differently from a fluid under rotation.
A more complicated question, however, has to do with
the effective value of $\varepsilon $ once the star is precessing,
because it is not clear how the structure of the star adjusts itself
under this circumstances.
The best way to calculate  $\varepsilon$ in a fully general relativistic
approach should take into account the crust and the precession of the star
simultaneously.
Such a model is obviously quite difficult to solve in a fully general
relativistic approach. Therefore, we expect that the actual value of
$\varepsilon$ may be lower than our present
estimates, but our results indicate that it is probably higher than  usually
assumed. Since values of $\varepsilon \sim 10^{-3} - 10^{-4}$
may be more adequate for ``slow'' rotating sources,
the presented results may be viewed as firm
upper limits.

\par We remark that other works have already considered
 pulsars as possible sources of gravitational waves, among
them  Press \& Thorne (1972), Zimmermann (1978, 1980), Zimmermann
\& Szedenits (1979), Thorne (1987), Barone et al. (1988),
Nelson et al. (1990) and Finn \& Shapiro (1990). In particular,
some of the authors among others, Zimmermann (1978, 1980),
Zimmermann \& Szedenits (1979), Barone et al. (1988) and
Nelson et al. (1990), addressed their studies to wobbling pulsars,
not considering, however, a general relativistic calculation of
$\varepsilon$ as we have done. We hope that our work can provide
quantitative results for a better evaluation of the problem.

\par Since the structure of a neutron star depends on
the adopted equation os state (hereafter EOS) we
should first specify the choices made on them. In the present work, we
performed
calculations using a new set of EOS's derived by Rufa et al. (1990),
which include the contribution of hyperons in the strongly interacting
matter. We have also calculated some models using the relatively
stiff Bethe-Johnson I (hereafter B\&JI) EOS in order to compare the
accuracy of our numerical code with previous computations.

\vfill\eject

\PN {\bf 2. Numerical Code}

\par\noindent  Our numerical code has been based on the approach developed
by BI76, Butterworth (1976) and FIP86.

\par The method developed by BI76 is a generalization
of Stoekly's (1965) work on rotating Newtonian polytropes.

\par To proceed, the distance element is first written in the form

\vskip 18pt
\PN $ds^2 = -e^{2\nu}c^2dt^2+r^2\sin ^2 \theta B^2 e^{-2\nu} (d\phi-\omega dt)
^2 + e^{2(\lambda-\nu)} (dr^2 + r^2 d\theta^2 ) \hfill  (1)$
\vskip 18pt

\PN with the metric functions $\nu , B, \omega$ and $\lambda$
independent of t and $\phi$.

\par Using the above metric in the Einstein equations we get differential
equations involving $\nu , B, \omega$ and $\lambda$. These equations
were solved numerically using the Newton-Raphson method, for a given
EOS and angular velocity $\Omega$.

\par The Newton-Raphson technique requires, however, a known solution
as a guess. We start the calculations from a spherical
solution as a guess. This solution is easily obtained by solving
the TOV equations for a given EOS and central density. As TOV equations
are written in the Schwarzchild coordinates, we
have to translate the results to the metric used here, which  was done
following Butterworth (1976). Thus, having the spherical solution
as a guess, we take a small
value of $\Omega$ and obtain a model for this $\Omega$. Models
for larger $\Omega$'s are obtained using previous models, with
lower values of $\Omega$. There is , however, an absolute
maximum $\Omega$ ($\Omega_{max}$) that
a uniformly rotating star can support. This $\Omega_{max}$ is the
Keplerian angular velocity ($\Omega_k$), namely, the angular
velocity of a particle in circular orbit at the equator. The maximum
rotation rate, however, depends on the EOS and on
the mass of the star (or central density).

\par In fact, although
the Keplerian angular velocity is an upper limit, Ipser \&
Lindblom (1989a, b) showed that the maximum rotation angular velocity
is actually
below the Keplerian angular velocity. They argued that gravitational
radiation instability implies that $\Omega_{lim} \simeq (0.86-0.94)\Omega_k$,
where the viscosity, included in their calculations,
works as a damping mechanism against instability, having therefore
an important role on the value of $\Omega_{lim}$. Other instability
analyses were performed by Lindblom (1992), Lindblom \& Mendell (1992)
and Weber \& Glendenning (1991), with similar results.
\par For a given EOS and fixed $\beta$ (the injection  energy
of a unit mass particle lowered from infinity to the star), it
is possible to get a sequence of models with $0< \Omega < \Omega_k$.
The choice of $\beta$, for given EOS, defines the mass of the star.
$\beta$ is related with the corresponding metric function in Eq.(1) by

\vskip 18pt

\PN $ \beta = e^{2\nu_p} \hfill  (2)$
\vskip 18pt

\PN where $\nu_p$ is the value of $\nu$ at the pole of the star.

\par The structure equation is given by the equation of hydrostatic
equilibrium which, for a uniformly rotating star, has a first integral

\vskip 18pt
\par\noindent $ h(p) = \ln {\beta^{1\over 2} e^{-\nu} \over
\sqrt{ 1- v^2/c^2}} \hfill  (3)$
\vskip 18pt

\PN where $h(p)$ is the comoving entalpy density,
$v=(\Omega-\omega)r\sin\theta B e^{-2\nu}$
is the velocity of the fluid relative to the zero angular momentum
observer and $c$ is the velocity of light.

\par On the other hand $h(p)$ can be obtained from the EOS, since we are
dealing
with zero-temperature matter, and the relation with pressure and energy density
is given by

\vskip 18pt
\par\noindent $ h(p) = \int_ 0^ p { dp^\prime \over e^\prime + p^\prime }
\hfill  (4)$
\vskip 18pt

\par Obtaining $h(p)$, from the above equation, we can generate a table
containing $h,p$ and $e$. Once $h(p)$ is determined from Eq.(3) we can
get $p$ and $e$ by interpolation. We, in particular, have used a cubic
spline interpolator.

\par Thus its possible to get $p(r,\theta)$ and $\rho(r,\theta)$
determining, therefore, the star's structure for a given EOS, mass (M)
and $\Omega$.

\par FIP86  obtained
several sequences for $ 0 < \Omega < \Omega_k $ for different EOS's in the
literature and fixed values of $\beta$. However, by this procedure, for
different values of $\Omega$ they obtained stars
with different number of baryons. Since we were interested in following
the variation of the star's structure as rotation
increases, we have chosen
to adjust $\beta$ in order to maintain fixed the total number of
baryons along the sequence $0 < \Omega < \Omega_k $.

\par In the method used here (Newton-Raphson), the equations
for $\nu$ , B, and $\omega $  are linearized. For
the resolution of these linearized equations, we transformed
them into difference equations on a finite grid in the
$(r,\mu)$ plane (where  $\mu\equiv\cos\theta$).
The metric functions $\nu , B$ and $\omega$ are even functions
of $\mu$ and  in this way the grid  covers the interval $0 < \mu < 1$.
For the angular coordinate the spokes are taken from the
Gauss-Legendre quadrature values $\mu_1 = 0, ... \mu_l$.
We particuraly choose $l=6$. The Gauss-Legendre quadrature technique
gives a good accuracy and , it is not necessary to include a large
number of spokes.
For the radial direction, we have used up to 120 spokes. In our
calculations convergence to one part in $10^3$ or $10^4$ was
required.

\par Our numerical code was checked by calculating some models with
B\&J I EOS and the same star parameters used by FIP86. The comparison
of the resulting properties is given in Table 1 and the agreement
between both set of calculations is quite good, giving confidence
to our numerical code.

\vfill\eject

\PN {\bf 3. Results}

\par\noindent As we mentioned before, our models were based on the EOS's by
Rufa et al. (1990). These EOS's represent a new approach concerning
the inclusion of hyperons in the dense nuclear matter. Those calculations
were performed in  the relativistic mean field approximation and depend
on two coupling constants of the theory, $g_{\omega\lambda}$ and
$g_{\sigma\lambda}$ . These constants are not well determined experimentally
and they were considered as parameters for the different set of EOS. In fact
Rufa et al. (1990) found that is even possible to have self-bound $\Lambda$
matter and, therefore, absolutely stable states at zero pressure. The
hyperon species are expected to be present above the nuclear saturation density
$\rho_{o}$ and they may be very important to understand the cooling history
of the stars since they may enhance the neutrino emissivity (see Pethick \&
Ravenhall 1992 and references therein).
In Fig. 1 we show their EOS's
calculated for different values of the coupling constants. For comparison,
we plotted also the B\&J I EOS. The relative importance of $\Lambda$'s
increases from EOS1 to EOS4 and the EOS becomes softer for decreasing
densities. Note that the EOS's are significantly different only for
$\rho > 6\times 10^{14} g cm^{-3}$. Table 2 gives the resultating
non-rotating maximum neutron star masses for those EOS's as well as
the respective values adopted for the vector and scalar coupling
constants. We can see, as it would be expected, that the masses decrease
as softer EOS's are used. Moreover, EOS4 allows a maximum mass of
only $1.26 {M_\odot}$. This value is smaller than $\sim \; 1.42 M_\odot$,
the masses of the neutron stars in the binary system PSR1913+16 and
thus such low values for the scalar and vector coupling constants are excluded.
Since EOS4 can be ruled out by these results, we have concentrated our
efforts on the remaining ones.

\par The properties of the stellar models are given in Table 3.
Figure 2 shows the gravitational deformation $\varepsilon$ as a function
of angular rotation velocity for the calculated set of models, including
also a B\&J I EOS model. As mentioned earlier in the introduction, the
values of $\varepsilon$ that we have obtained may be an upper limit,
since an actual modelling should involve the detailed structure of
the neutron star, in particular, its crust and the precession.

\par Let us now estimate the value of the amplitude of the gravitational
waves, $ h$, for a wobbling pulsar considering the models that we have studied.
The amplitude can be given by $ h=(16\pi GF/c^3\Omega^2)^{1/2}$
(see, e.g., Zimmermann 1978), where $F=L_{GW}/4\pi r^2$ is the flux
with $L_{GW}$ being the gravitational wave luminosity and $r$ is
the distance to the star. $L_{GW}$ can be taken from, e.g., Zimmermann (1980)
or Shapiro \& Teukolsky (1983) who provide us the gravitational
wave luminosity for a wobbling star. Thus, we can write

\vskip 18pt
\par\noindent $h\simeq 4\times 10^{-26}\varepsilon\theta_w
\biggl( {I_{zz}\over 10^{45}g.cm^2} \biggr)
\Omega^2 r_{kpc}^{-1} \hfill  (5)$

\vskip 18pt

\PN where: $\varepsilon$ is the gravitational
deformation, $\theta_w$ is the wobble angle, $I_{zz}$ is the
moment of inertia with respect to the rotation axis,
$\Omega$ is the angular velocity in rad/s and $r_{kpc}$ is the
distance to the star in kpc. It is worth mentioning that the above
equation for $h$ gives us only a rough idea (probably not better than the
order-of-magnitude)
of the true amplitude, due to approximations done in its
derivation. The equation for $h$ is, in fact, derived for a rigid
newtonian object rotating free of external torques in the
standard quadrupole moment formalism, and it is not completely
appropriate, therefore, for strong gravitational fields.

\par In Table 4 we present the results of our calculations assuming
$\theta_w = 10^{-5}$, for the stars modeled in Table 3, at a distance
$r_{kpc}= 20$, rotating at the keplerian angular velocity. We
have obtained $h \sim 10^{-25}$ which might be detected
when, for example, the Caltech-MIT LIGO antenna  (see, e.g., Thorne 1987 and
Finn \& Shapiro 1990) or the VIRGO antenna (see, e.g., Giazzoto 1989 and
Bradaschia et al 1990) become operative but only if the
wobble motion is continuously excited at least for several months.
Otherwise, ``spike'' (impulsive bursts) with durations of $\sim \; 1$ s
may result if the wobble is excited by some catastrophic phenomenon
(e.g. phase transitions) but damps out on a quadrupole emission timescale
(see, e.g., de Araujo et al 1994). Similar values for $h$, but considering
triaxial deformation of the pulsar,
were obtained by Finn \& Shapiro (1990).

\vfill\eject

\PN {\bf 4. Conclusions}

\par\noindent The present calculations of rotating neutron stars were
performed using a new set of EOS's, including the presence
of $\Lambda$'s in the strongly interacting matter (Rufa et al. 1990).
The resulting maximum masses for the neutron star configurations
give bounds on the values of the scalar and vector coupling
contants, suggesting $g_{\omega\lambda}$, $g_{\sigma\lambda}$ $ ^>_\sim$  0.43,
otherwise too soft EOS's are obtained.

\par If we take stars with the same mass for the Rufa et al. EOS's,
it is not possible to distinguish them through their rotational
behavior. In other words, the existence of fast millisecond pulsars
is not at odds with exotic possibilities like self-bound $\Lambda$
matter. The rotational properties of neutron stars were investigated
for configurations having a constant number of baryons, condition
required for obtaning consistent theoretical sequences which may be
considered as an evolutionary path for decelerating objects.

\par There is a potentially important application of this paper
related to the fully general relativistic
values derived for the gravitational deformations $\varepsilon$,
which may be two up three orders of magnitude (at best) larger than is usually
entertained. We have shown that these values of $\varepsilon$, at
relatively small wobble angles, and observed rotation rates, might allow the
detection of sources located in the Galaxy when the advanced
generation of antennas becomes operative.

\par Finally, it is worth mentioning that the calculations of
$\varepsilon$ may be improved, since it is possible, in principle,
to calculate it, in a full general relativistic approach, including
in detail, e.g., the crust of the neutron star and the precession.
In doing this we could have a lower value of $\varepsilon$, as compared with
the values that we have obtained in the present work.
But these new values would still be higher than  usually
considered in the literature, because these last values are
adequate for ``slow'' radio pulsars.
\par It is our aim in the future to perform a more detailed calculation
of $\varepsilon$ taking in account the crust of the neutron star and
the precession with the use of our general relativity numerical code.

\vfill\eject

\PN{\it Acknowledgements.} We would like to thank the Brazilian agencies
FAPESP, CNPq and CAPES for support. We also would like to thank Dr. M. Rufa
for giving us the Frankfurt EOS's in table form. JCNA would like to thank
Dr. J.C. Miller and Dra. M. Colpi for useful comments and discussions.
The final version of this paper was written while
JCNA was visiting the International School for Advanced Studies
(SISSA/ISAS), he would like to thank the hospitality received there,
in particular, from the head of the Astrophysics Sector, Prof. D.W. Sciama.
Finally, we would like to thank our referee, Dr. E. Gourgoulhon, for
his careful reading of the paper and for useful comments and suggestions
that greatly improved the final version of our paper. The calculations
have been made with the VAX/8530 computer purchased by FAPESP.

\vfill\eject

\PN {\bf Appendix: The calculation of $\varepsilon$}

\par The parameter $\varepsilon$ is defined as:

\PN $ \varepsilon = (I_{zz}-I_{xx})/I_{zz} \hfill (A1)$

\PN therefore, it is necessary to calculate the moments of inertia to
obtain it.
Before going into the details of our calculation
of the moments of inertia it is important to characterize
the gravitational wave amplitude $ h$.
Strictly speaking, due to the fact that the formula to
calculate $h$ is newtonian (or, at best, pos-newtonian)
one might use, to be consistent, a newtonian (or pos-newtonian)
calculation of the moments of inertia.

\par Even if one had a definite formula for $h$
in a fully general relativity version its calculation would
be problematic, due to the fact that it is not
possible to define $I_{xx}$ (or $I_{yy}$) invariantly,
as done for the $I_{zz}$ through the
definition:

\PN $ I_{zz}= J/\Omega \hfill (A2)$

\par\noindent (where $J$ is the angular momentum
and $\Omega$ is the angular velocity).

\par It should be kept in mind that the formula
of $h$ gives us only an order-of-magnitude estimate
of the true amplitude, due to the various approximations in
its derivation. It is worth mentioning that a newtonian calculation
of the $I_{ij}$'s should be, in principle, acceptable
due to the fact that the equation used to calculate the gravitational wave
amplitude is also derived for a rigid newtonian object.
To calculate the $I_{ij}$'s using the newtonian theory
would be, on other hand, misleading, because we use fully general
relativity theory to model our rapidly rotating stars.
Thus, we decided to calculate all the $I_{ij}$'s
through the formula:

\PN $ I_{ij} = \int (r^2\delta_{ij} -x_ix_j)dM \hfill (A3)$

\par\noindent where r is the radial coordinate of the metric used
and the $x_i$'s are their projections, and $dM$ is the gravitational
mass element, namely

$$\eqalignno{ dM &=(-T^0_0+T^1_1+T^2_2+T^3_3)\sqrt{-g}\,dr\,d\theta\, d\phi
& (A4) \cr &=\biggl\{Be^{2\lambda-2\nu}\biggl[2p+{(e+p)(1+v^2)\over 1-v^2}
\biggr]+2\,r\sin\theta\,\omega \, B^2e^{2\lambda-4\nu}\,
{(e+p)v\over 1-v^2}\biggl\}\,r^2\sin\theta \,dr\,d\theta\, d\phi \;\;\, ,\cr
}$$

\PN where $\lambda$, $\nu$, B and $\omega$ are metric functions,
and $v$, $p$ and $e$ are the velocity, the pressure
and the energy density, respectively.

\par This definition, is in fact, a newtonian-like way to
calculate the moments of inertia, and therefore
not invariantly defined, although it takes into account
how the matter is distribuited throughout the star structure.

\par We point out that our formula to calculate
the $I_{ij}$'s produces $I_{zz} \not= J/\Omega$, although,
for the models that we have studied, the difference is
$\leq 10\%$ throughout the whole sequence, i.e., $ 0< \Omega< \Omega_k$.

\par Finally we note that, although
the definition of the $I_{ij}$'s is not unique, the
values for $\varepsilon$ should not change strongly if one adopts
another definition, because, this quantity is an implicit monotonic
function of the eccentricity after all.

\vfill\eject

\PN {\bf References}

\PN Barone, F., Milano, L., Pinto, I., Russo, G., 1988, A\&A, 203, 322

\PN Bonazzola, S., Gourgoulhon, E., Salgado, M., Mark, J.A., 1993, A\&A,
278, 421

\PN Bradaschia, C. et al. 1990, Nucl. Instr. Meth. A, 289, 518

\PN Butterworth, E.M., 1976, ApJ, 204, 561

\PN Butterworth, E.M., Ipser, J.R., 1976, ApJ, 204, 200 (BI76)

\PN Cook, G.B., Shapiro, S.L., Teukolsky, S.A., 1992, ApJ, 398, 203

\PN Cook, G.B., Shapiro, S.L., Teukolsky, S.A., 1994a, ApJ, 422, 227

\PN Cook, G.B., Shapiro, S.L., Teukolsky, S.A., 1994b, ApJ, 424, 823

\PN de Araujo, J.C.N., de Freitas Pacheco, J.A., Horvath, J.E.,
Cattani, M., 1994, MNRAS, 271, L31

\PN Eriguchi, Y., Hachisu, I., Nomoto, K., 1994, MNRAS, 266, 179

\PN Ipser, J.R., Lindblom, L., 1989a, Phys. Rev. Letters, 62, 2777

\PN Ipser, J.R., Lindblom, L., 1989b, Phys. Rev. Letters, 63, 1327

\PN Finn, L.S., Shapiro, S.L., 1990, ApJ, 359, 444

\PN Friedman, J.L., Ipser, J.R., Parker, L., 1986, ApJ, 304, 115 (FIP86)

\PN Giazzoto, A. 1989, Phys. Rep., 182, 365

\PN Komatsu, H., Eriguchi, Y., Hachisu, I., 1989a, MNRAS, 237, 355

\PN Komatsu, H., Eriguchi, Y., Hachisu, I., 1989b, MNRAS, 239, 153

\PN Lattimer, J.M., Prakash, M., Masak, D., Yahil, A., 1990, ApJ,
355, 241

\PN Lindblom, L. 1992, Instabilities in Rotating Neutron Stars.
In: Pines, D., Tamagaki, R., Tsuruta, S. (eds.) The Structure and
Evolution of Neutron Stars. Addison-Wesley, New York, p. 122

\PN Lindblom, L., Mendell, G., 1992, Superfluid Effects on the Stability
of Rotating Newtonian Stars. In: Pines, D., Tamagaki, R., Tsuruta, S. (eds.)
The Structure and Evolution of Neutron Stars. Addison-Wesley, New York, p. 227

\PN Nelson, R.W., Finn, L.S., Wasserman, I. 1990, ApJ, 348, 226

\PN Pethick, C.J., Ravenhall, D.G. 1992, Phil. Trans., 341, 17

\PN Pines, D., Shaham, J. 1974, Nature, 248, 489

\PN Press, W.H., Thorne, K.S. 1972, ARA\&A, 10, 335

\PN Rufa, M., Schaffner, J., Maruhn, H. Stoekler, H., Greiner, W.
1990, Phys. Rev. C, 42, 2469

\PN Salgado, M., Bonazzola, S., Gourgoulhon, E., Hansel, P. 1994, A\&A,
291, 155

\PN Shapiro, S.L., Teukolsky, S.A., 1983. Black Holes, White Dwarfs
and Neutron Stars: The Physics of Compact Objects. Wiley Interscience,
New York

\PN Stoeckly, R. 1965 , ApJ, 142, 208

\PN Thorne, K.S. 1987, Gravitational Radiation. In: Hawking, S., Israel, W.
(eds.) 300 Years of Gravitation. Cambridge University Press, London, p. 330

\PN Weber, F., Glendenning, N.K. 1991, Z. Phys. A, 339, 211

\PN Zimmermann, M. 1978, Nature, 271, 524

\PN Zimmermann, M. 1980, Phys. Rev. D, 21, 891

\PN Zimmermann, M., Szedenits, E. 1979, Phys. Rev. D, 20, 351

\vfill\eject

\centerline {\bf Figure Captions}
\vskip 24pt

\par\noindent {\bf Figure 1.} Frankfurt and B\&J I EOS's (see the text for
details)
\vskip 18pt
\par\noindent  {\bf Figure 2.} Gravitational deformation
$\varepsilon$ as a function of $\Omega$ for the models 1, 2, 3
\par\noindent and B\&J I.

\vfill\eject

\baselineskip=18pt

\par\noindent {\bf Table 1.} Comparison between our calculations and
FIP's calculations (in parenthesis). Where: $\beta$ - injection energy,
$\Omega$ - angular velocity, $\rho_c$ - central density;
$M$ - gravitational mass,
$R_{eq}$ - equatorial radius; $\omega_c/\Omega$ -
central dragging in units of angular velocity; T/W - ratio of
rotational energy to gravitational energy, $v_{eq}/c$ - velocity of
a comoving observer at the equator relative to a locally nonrotating observer,
and $\epsilon_c=\sqrt{1-R_p^2/R_{eq}^2}$- eccentricity (where $R_p$ is
the polar radius).

$$\vbox{\halign{\hfil#\hfil&\qquad\hfil#\hfil&\qquad\hfil#
\hfil&\qquad\hfil#\hfil&\qquad\hfil#\hfil&
\qquad\hfil#\hfil&\qquad\hfil#\hfil&\qquad\hfil#\hfil&
\qquad\hfil#\hfil\cr
\noalign{\hrule}
\ \cr

$\beta$ & $\Omega$ & $\rho_c$ & $M/M_{\odot}$ & $R_{eq}$ &
$\omega_c/\Omega$ & $T/W$ & $v_{eq}/c$ & $\epsilon_c$ \cr
$ $ & $(rad/s)$ &  $(10^{15}g/cm^3)$  &  $ $  &  $(km)$  &  $ $ & $ $ & $ $\cr

\ \cr
\noalign{\hrule}
\ \cr

$ $ & $ 0 $ & $ 1.00 $ & $1.34$ & $12.3$ & $-$ & $-$ & $-$ & $-$\cr
$ $ & $(0)$ & $(1.00)$ & $(1.32)$ & $(12.1)$ & $-$ & $-$ & $-$ & $-$\cr

$ $&$3000$&$0.95$&$1.33$&$12.6$&$0.43$&$0.017$&$0.14$&$0.32$\cr
$0.671$&$(3000)$&$(0.95)$&$(1.31)$&$(12.5)$&$(0.42)$&$(0.017)$&$(0.13)$
&$(0.27)$\cr
$(0.676)$&$4030^\ast$&$0.91$&$1.32$&$13.3$&$0.42$&$0.033$&$0.19$&$0.49$\cr
$ $&$(4030)$&$(0.91)$&$(1.30)$&$(13.3)$&$(0.42)$&$(0.034)$&$(0.18)$&$(0.47)$\cr

$ $&$5746^\dagger$&$0.79$&$1.31$&$16.2$&$0.41$&$0.087$&$0.34$&$0.73$\cr
$ $&$(5700)$&$(0.77)$&$(1.29)$&$(16.9)$&$(0.40)$&$(0.093)$&$(0.32)$&$(0.74)$\cr

\ \cr
\noalign{\hrule}}}$$
\par\noindent $^\ast$ Angular velocity of the fatest pulsar known
\par\noindent $^\dagger$ Keplerian angular velocity

\vfill\eject

\par\noindent {\bf Table 2.} Non-rotating maximum neutron star masses
for the Frankfurt EOS's as well the respective values of the
coupling constants $g_{\omega\lambda}$ and $g_{\sigma\lambda}$.

$$\vbox{\halign{
\hfil#\hfil&\qquad
\hfil#\hfil&\qquad
\hfil#\hfil&\qquad
\hfil#\hfil  \cr
\noalign{\hrule}
\ \cr

$EOS$ & $g_{\sigma\lambda}$ & $g_{\omega\lambda}$ & $M_{max}/M_{\odot}$ \cr

\ \cr
\noalign{\hrule}
\ \cr

$1^\ast$   &   $  - $   &  $ -  $  &  $2.17$ \cr
$2     $   &   $0.80$   &  $0.70$  &  $1.98$ \cr
$3     $   &   $0.43$   &  $0.43$  &  $1.54$ \cr
$4     $   &   $0.0 $   &  $0.10$  &  $1.26$ \cr

\ \cr
\noalign{\hrule}}}$$
\centerline {$^\ast$ No $\Lambda$'s are present}

\vfill\eject

\par\noindent {\bf Table 3.} Results of our calculations for
the Frankfurt EOS's and B\&J I EOS. Where $M_o$ is the baryon
mass that is held constant ( within $\pm 1 \%$) through
the calculations of each model.

$$\vbox{\halign{\hfil#\hfil&\quad\hfil#\hfil&\quad\hfil#
\hfil&\quad\hfil#\hfil&\qquad\hfil#\hfil&
\qquad\hfil#\hfil&\qquad\hfil#\hfil&\qquad\hfil#\hfil&
\qquad\hfil#\hfil\cr
\noalign{\hrule}
\ \cr

$MODEL$ & $EOS$ & $M_o/M_{\odot}$ & $M/M_{\odot}$ &
$\Omega$ & $\varepsilon$ & $\epsilon_c$ & $I_{xx} \qquad I_{zz}$ &
$R_{eq}$ \cr

$ $ & $ $ & $ $ & $ $ & $(rad/s)$ & $ $ & $ $ & $(10^{45}g.cm^2)$ &
$(km)$ \cr

\ \cr
\noalign{\hrule}
\ \cr

$ $ & $1,\;2 $ & $1.06$ & $1.01$ & $3000$ & $0.12$ & $0.50$ & $0.78\qquad 0.88$
& $16.0 $\cr

$I$ & $and\;3 $ & $1.08$ & $1.02$ & $3950^\dagger$ & $0.22$ & $0.75$ &
$0.83\qquad 1.06$ & $19.8 $\cr

\ \cr
\noalign{\hrule}
\ \cr

$ $ & $1,\;2 $ & $1.45$ & $1.57$ & $4030$ & $0.15$ & $0.52$ & $1.01\qquad 1.19$
& $15.8$\cr

$II$ & $and\;3 $ & $1.46$ & $1.57$ & $4900^\dagger$ & $0.24$ & $0.75$ &
$1.10\qquad 1.45$ & $19.2 $\cr

\ \cr
\noalign{\hrule}
\ \cr

$ $ & $1$ & $2.02$ & $1.82$ & $4030$ & $0.10$ & $0.43$ & $1.01\qquad 1.12$
& $14.4$\cr

$III$ & $and\;2$ & $2.02$ & $1.84$ & $5696^\dagger$ & $0.23$ & $0.67$ &
$1.16\qquad 1.50$ & $17.2$\cr

\ \cr
\noalign{\hrule}
\ \cr

$ $ & $ $ & $1.59$ & $1.45$ & $4030$ & $0.089$ & $0.38$ & $0.64\qquad 0.70$
& $12.8$\cr

$B\&J\;I $ & $B\&J\;I $ & $1.60$ & $1.47$ & $6200^\dagger$ & $0.25$ & $0.75$
& $0.75\qquad 0.99$ & $16.6$\cr

\ \cr
\noalign{\hrule}}}$$
\par\noindent $^\dagger$ Keplerian angular velocity

\vfill\eject

\par\noindent {\bf Table 4.} Values of h using Eq. 5 for the models I-III and
B\&JI, for the Keplerian angular velocity, taking pulsars at $r_{kpc}=20$
and with $\theta_w=10^{-5}$.

$$\vbox{\halign{
\hfil#\hfil&\qquad
\hfil#\hfil  \cr
\noalign{\hrule}
\ \cr

$MODEL$ & $ h\; (\times 10^{-26})$ \cr

\ \cr
\noalign{\hrule}
\ \cr

$I       $   &   $ 7 $   \cr
$II      $   &   $ 17 $   \cr
$III     $   &   $ 22$    \cr
$B\&J\; I$   &   $ 19$    \cr

\ \cr
\noalign{\hrule}}}$$

\bye